\documentstyle[12pt]{article}              

\title{{\hfill{\small{BGU PH-97/01}}}\protect\\ 
Inclusive spectra of hadrons
 created by color tube fission 1. Probability of tube fission}
\author{E.V.Gedalin\thanks{E-mail address: gedal@bgumail.bgu.ac.il}}
\date{Department of Physics \\
 Ben-Gurion University of the Negev\\
Beer-Sheva, 84105,  Israel}
\begin{document}
\maketitle
\ \ \ 
\newpage
\begin{abstract}
The probability of color tube fission that includes the tube surface small
oscillation corrections is obtained with pre-exponential factor accuracy  on
the basis of previously constructed color tube model. Using these
expressions the probability of the tube fission in $n$ point is obtained 
that is the basis for calculation of inclusive spectra of produced hadrons.
\end{abstract} 

\pagebreak 
\section{ Introduction}

\bigskip
\ indent In the previous papers [1-4] the properties of color tubes and
their fission are considered on the basis of a
semi-phenomenological effective lagrangian of (1+1) - dimensional
theory of two scalar fields $r$ and $\phi$ describing the
simultaneous change of color tube chromoelectric field and
radius [1]. From the point of view of our theory a color tube is a
soliton kink ($K$)+antikink ($\bar{K}$)- like solutions of field
equations. The color charges are arranged in walls $K$ and
$\bar{K}$ that are moving apart in opposite directions with large
momenta $P_K$  and $P_{\bar{K}}$. The color tube is a metastable state 
and it is "hadronized" by fission to primary hadrons, which decay
into observable particles.

   We are interested in inclusive spectra (IS) of primary hadrons
created by fission of a tube with small surface oscillations. Our 
approach to the tube fragmentation into hadrons is close to Artru- 
Menessier model [5](as well as other models [6-9] based on Artru - 
Menessier approach). From the point of view of our effective 1+1 
dimensional theory a color tube is a kink and an antikink ($K$ and 
$\bar{K}$) that are moving apart in opposite directions and between which 
the color tube is formed. The breaking of the tube is a transition from 
the state of a pair ($K, \bar{K}$) to a state($K, \bar{K}_1, K_1, 
\bar{K}$)that is degenerate with it and such that, between extra pair  
($K_1, \bar{K}_1$) that have been formed inside the tube, we have a true 
vacuum state (the color field is equal to zero).

 At each stage of hadronization process the tube splits into pieces 
 with arbitrary masses as far as the distance between the $K$ and 
 $\bar{K}$ walls of piece becomes of the order of thickness of the wall.
  This causes  growth of kink mass and 
rapid decrease  of the probability of tube fission and therefore 
the hadronization process stops. The produced pieces form the primary hadrons 
that can be attributed to stable hadrons (pions and kaons) as well as 
hadron resonances. 

 We limit
ourselves by particles with light cone momentum $P_+=E+P$ small in comparison
with the initial light cone momentum of the tube $P_{+0}$.
Generally the procedure of IS calculation is the following. We must
integrate over the probability of tube fission at $n(n\geq2)$ space-time points
that formed the pieces  with given light cone momenta over all allowed
positions of fission space-time points [5,9,10].

   The basic quantity for calculation is the probability of tube fission as a
function  of fission point $P=P(\eta,\tau)$ where $\eta$ and $\tau$ are the
dimensionless space  and time coordinates of fission point. We shell use the
probability $P$ that includes the tube surface small oscillation corrections.

   In our tube model there are two kinds of surface oscillations with 
 large and small quanta masses that play different roles in the tube 
 formation and subsequent evolution [1-4]. The large -mass oscillations 
 play essential role in the formation of color tube [2]: they lead 
 to the breaking of the tube ( with the
probability $P\approx1$) that just begins grow.
\footnote [1]{Perhaps small jets that were observed in high energy 
hadroproduction experiments are mainly produced by this way. 
Hadrons in such a jets are created by fission of sufficiently 
short pieces produced in early stage of tube
formation via catastrophic breaking of growing tube.}
Only small amplitude of $a$ small mass  oscillations remain in created pieces
[2]. Fission probability of such a tube is small and accordingly 
its evolution time is large. The tube grows  lengthwise and its 
length become much larger then radius.
 
   Small mass oscillations produce corrections to exponential factor of 
   fission probability of long tube that are proportional to oscillation 
   amplitude squared
\begin{equation}
 P(\eta,\tau)\sim\exp-[\pi M^2/\rho+a^2 D(s)].
\end{equation}
where $D(s)$ is the function which explicit form depends on initial shape
of color tube. Here $s=\tau^2-\eta^2$ and $M$ and $\rho$ are the 
dimensionless kink mass and tube tension respectively.

   For IS calculation we need to more accurately calculate the fission
probability including the pre-exponential factor corrections. Therefore 
before
going to calculate IS we
discuss (in section 2) the general approach to pre-exponential factor
calculation for tunneling transition process from false vacuum + smal
 classical field to true vacuum and 
 calculate a fission probability
including the tube surface oscillations corrections with pre-exponential
factor accuracy (section 3). In Section 4 we obtain a general
formula for tube breaking in $n$ different space-time points that is
leading to production of $m$ pieces with lengths $l_1, l_2, ...l_m$
respectively.  Finally, we summarize the results in Conclusions.

\bigskip
\section{The probability of tunneling transition \protect\\
with pre-exponential accuracy}
 \bigskip
\indent Here we shell explore the pre-exponential factor in the
probability $P$ (per unit time and per unit length) of transition
false vacuum + classical field $\rightarrow$ true vacuum in Minkowski
space. 

Following the general approach of Ref.[12] we consider the
Lagrangian 
\begin{equation}
L(A)=(1/2)({\partial_\mu}A)^2 - U(A),
\end{equation}
where $A(x,t)$ is the field variable, $\mu=0,1$;
$\partial_0=\partial/\partial t$, $\partial_1=\partial/\partial x$;
the metric is $(+,-)$. We denote the initial state false vacuum and
classical field variables respectively by $A_1$ and $A_c$ and
final state true vacuum field by $A_0$.

The tunneling transition in which we are interested takes place
from the state $A_i=A_1+A-c$ into such state $A_f$ that for $x<x_L$
and $x>x_R$ we have $A=A_i$ and for $x_L<x<x_R$ the field $A=A_0$.
We assume that $A_c$ is the field of small classical oscillations
and express it in the form
\begin{equation}
A_c =\sum a_n\Theta_n(x_0,t_0),
\end{equation}
where $a_n\gg1$ and $\Theta_n$ are the amplitudes and modes of
normal oscillations, respectively.

As it was shown in Ref.[12] the "most probable escape path"
(MPEP) in the function space connecting from $A_i$ to $A_f$ can be
described by $A(x,t)= f(x,\lambda(t))$ where $\lambda(t)$ is the
function of time. To obtain the solution of tunneling problem the
pair $(\lambda,\dot{\lambda}={\partial_0}\lambda)$ must be considered
as the dynamical variables and the field functional
integral should be treated as a path integral over functions $\lambda$.

Let us consider first the case when all $a_n=0$. As it was shown
in Ref.[2] the proper choice for collective variable
$f(x,\lambda)$ is the kink-antikink ($K,\bar{K}$) configuration
solution of field equations  with $K$ and $\bar{K}$ located at
$\Lambda_K$ and $\Lambda_{\bar{K}}$, moving with velocities 
$\dot{\Lambda}_K$ and $\dot{\Lambda}_{\bar{K}}$ respectively [2]
\begin{equation}
f(x,\lambda)=A_{K,\bar{K}}\nonumber \\
=K(\frac{x-\Lambda_K}
{(1-{\dot{\Lambda}}_K^2)^{1/2}})+
{\bar{K}}(\frac{x-\Lambda_{\bar{K}}}
{(1-{\dot{\Lambda}}_{\bar{K}}^2)^{1/2}})+C.\label{akk}
\end{equation}
It must be noted that $A_{K,\bar{K}}=0$ inside the interval
$(\Lambda_K,\Lambda_{\bar{K}})$ and $A_{K,\bar{K}}=1$ far away from this region.  
It is obvious that the MPEP parameter $\lambda$ coincide with
$\Lambda_{\bar{K}}$ at $\lambda>x_0$ and with $\Lambda_K$ at $\lambda<x_0$
where $x_0$ is the center of the $K,\bar{K}$ pair. Thus the transition 
amplitude for tunneling process can be written in the form
\begin{equation}
<A_f| A_i>=\int {\cal{D}} \lambda (t)
\exp [-i \int dt L_{eff}(\lambda(t))] \label{path}
\end{equation}
where
\begin{equation}
L_{eff}=\int dx (L[A_{K\bar{K}}]-L[A_i])
\end{equation}
is the effective Lagrangian of dynamical variable $\lambda(t)$ that
describes the tunneling process.

It is easy to verify that in the thin wall approximation (i.e. when 
the sizes of $K$ and $\bar{K}$ are much less then $2M/\rho$) 
for $L_{eff}$ we obtain
\begin{equation}
L_{eff}=-M[(1-{\dot{\Lambda}_K}^2)^{1/2}
+(1-{\dot{\Lambda}_{\bar{K}}}^2)^{1/2}]+\rho(\Lambda_{\bar{K}}-
\Lambda_K)\label{lag}
\end{equation}
where $M$ is  the mass of $K$ and $\bar{K}$,
\begin{equation}
M=\int dx (\partial_{\lambda}A_{K\bar{K}})^2\mid_{\lambda =
\Lambda_K;\Lambda_{\bar{K}}}
\end{equation}
The first term in (7) is the Lagrangian of the free motion of $K$
and $\bar{K}$ with velocities $\dot{\Lambda}_K$ and
$\dot{\Lambda}_{\bar{K}}$, respectively, and the second term is the
contribution to the volume energy of the region between $K$ and
$\bar{K}$ from $U(A_{K,\bar{K}})-U(A_i)$, $\rho$ is the (constant) 
energy density per unit length.
 Now we proceed to calculate the $\lambda(t)$ path-integral in
the quadratic approximation. According to the semiclassical ideology
[12-21] the quadratic approximation for tunneling amplitude path-integral
(\ref{path}) can be obtained by expanding of the effective action around the
MPEP that is classical solution of equation of motion for 
$\lambda_c(t)$ for imaginary time $\tau=-it$ [18]. Thus we have
\begin{equation}
\lambda(t)=\lambda_c(\tau)+\lambda,
\end{equation}
where $\lambda_c$ obeys the equations
\begin{eqnarray}
-M\frac{d}{d\tau}({\dot{\Lambda}_K}/(1+{\dot{\Lambda}_K}^2)^{1/2})
=\rho,\nonumber \\
M(\frac{d}{d\tau}(\dot{\Lambda_{\bar{K}}}/(1+
{\dot{\Lambda_{\bar{K}}}}^2)^{1/2}))=\rho
\end{eqnarray}
with the initial conditions [2]
\begin{eqnarray}
\Lambda_K=\Lambda_{0K}=M/\rho, \\ \dot{\Lambda_K}=0,\\
\Lambda_{\bar{K}}=\Lambda_{0\bar{K}}=-M/\rho,\\
\dot{\Lambda_{\bar{K}}}=0, 
\end{eqnarray}
 at $\tau=0$.

Now after simple calculations  in quadratic approximation  for amplitude
$|<A_f|A_i>|$ we obtain
\begin{equation}
|<A_f|A_i>|=|I|\exp(-S_c) ,
\end{equation}
where $S_c$ is the imaginary part of "classical" action for MPEP
$\lambda_c(\tau)$
\begin{equation}
S_c=\pi M^2 / 2\rho,
\end{equation}
and
\begin{equation}
I=\int {\cal {D}} \sigma
\exp(-(M/2) \int_{\rho\tau_K / M}^{\rho\tau_{\bar{K}} / M} dz (1- z^2)^{1/2}
(\partial \sigma / \partial z)^2), 
\end{equation}
with boundary conditions for $\sigma$
\begin{equation}
\sigma(\rho\tau_K/M)=\sigma(\rho\tau_{\bar{K}}/M)=0.
\end{equation}
The time of the motion $K$ and $\bar{K}$ under barrier
$\tau_K$ and $\tau_{\bar{K}}$ are given by the relations [2]
\begin{eqnarray}
\tau_K\approx M/\rho;\, \tau_{\bar{K}}\approx-M/\rho;
\end{eqnarray}
and  we obtain for $I$ the following simple
expression
\begin{equation}
I=\int {\cal{D}} \sigma \exp(-\int_{-1}^{1} dx
(1-x^2)^{-1/2}\sigma H \sigma)
\end{equation}
where
\begin{equation}
H=(1-x^2)^{1/2}(d/dx)(1-x^2)^{1/2}(d/dx)
\end{equation}

The path integral $I$ can be easily calculated by  generalized
$\zeta$-function method [10,18] and after extracting zero mode 
contribution [12,13,19] we obtain 
\begin{equation}
 |I|=(V \rho / 2\pi)^{1/2}
\end{equation}
where V is the volume of space-time.

Thus for transition probability $P$ we obtain the following expression
\begin{equation}
P=(\rho / 2\pi)\exp(-\pi M^2 / 2\rho).\label{prob}
\end{equation}

The above expression is closely related to the result previously
obtained by Voloshin [21], who calculated transition
probability by using the "bubble" effective Lagrangian in
Euclidean space. As we have seen the same expression can be
derived also directly in Minkowski space although  the Minkowski
$L_{eff}$ differs from Euclidean $L_{eff}^E$.

Now we will show that the $L_{eff}$ have the form (22) for wide
class of Lagrangians allowing a true and false vacuum states and
$K$ and $\bar{K}$-like instanton solutions  of field equations 
that connect the true vacuum to the false vacuum.

Let us consider the nonlinear lagrangian of fields $A_l$
($l$=1,2,...$N$)that has the form
\begin{equation}
L=(1/2)F_{n,m}(A)(\partial_{\mu}A_n)(\partial_{\mu}A_m) + U(A), 
\label{lag1}
\end{equation}
where the matrix $F_{m,n}$ and potential $U(A)$ are the functions 
of fields $A_l$.

We assume existence of the extrema of $U(A)$ on sets of fields
${A^0}_n$ and ${A^f}_n$ that correspond to true and false vacua
respectively and $K$ - and $\bar{K}$ -type solutions of field
equations. 

Without any loss of generality we can assume that $F_{m,n}$ is a diagonal
matrix. Then functional integral measure for lagrangian (\ref{lag1}) has the
form 
\begin{equation}
{\cal{D}}[A] = \prod_{n} {\cal{D}} A_n \Delta_F^{1/2}
\end{equation}
where $\Delta_F$ is the determinant of matrix $F_{m,n}$.

Let us introduce new field variables $\varphi_n$ as follows
\begin{equation}
\varphi_n=\varphi_n(A) \label{nvar}
\end{equation}
in such a way that the Jacobian of this variable transformation 
is proportional to $\Delta_F^{-1/2}$. In these new field variables 
the functional integration measure has a very simple form
\begin{equation}
{\cal{D}}[\varphi]=\prod{d \varphi_n}
\end{equation}
and for the Lagrangian we obtain the following expression
\begin{equation}
L=\sum_{m,n}F_{m,n}(\varphi)(\partial_{\mu}\varphi_m)(\partial_{mu}\varphi_n)
-U(A(\varphi)).\label{lag2}
\end{equation}
where
\begin{equation}
F_{m,n}(\varphi)=\sum_{l,k}F_{l,k}(A)(\partial{A_l}/\partial{\varphi_m})
(\partial{A_k}/\partial{\varphi_n})
\end{equation}

From our assumption (\ref{nvar}) it follows that the MPEP in the 
$\varphi$-function
space is the $K-\bar{K}$-type $\varphi_n^{K \bar{K}}$ solution of the
field equations with the same $\lambda(t)$ for all $n$. 
\footnote[2]{Here we are
restricted by the case of one parameter $\lambda(t)$. It is clear 
that if $\lambda(t)$ for any $A_n$ (and, consequently, for $\varphi_n$) 
differs from the others only inside of $K$ or $\bar{K}$ walls and 
in the used thin wall approximation we can restrict ourselves by one 
path integral variable $\lambda(t)$. It is quite trivial to see that 
in the case of two or more different MPEP's the transition
probability is the sum of contributions of all MPEP's}

 The "Lorentz invariance" tells us that the $\varphi_n^{K\bar{K}}$ are dependent
only on variable $(x-\lambda)(1-{\dot\lambda}^2)^{-1/2}$. Now substituting in
~(\ref{lag2}) the new MPEP  $\varphi^{K\bar{K}}$ and passing to path 
integral over $\lambda$ we obtain for $L_{eff}$ in the thin wall approximation 
the same expression (\ref{lag}) with new definitions of $M$ and $\rho$
\begin{eqnarray}
M= \int dx \sum  F_{kl} [\partial {\varphi_k}^{K \bar{K}}/\partial x]
[\partial{\varphi_l}^{K\bar{K}}/\partial x],\\
\rho=(\Lambda_{\bar{K}}-
\Lambda_K)^{-1}\int dx [U(\varphi^{K\bar{K}})-U(\varphi^{f})]
\end{eqnarray}
where $\varphi^{f}$ is set of $\varphi$ -fields that correspond to set of
$A_{f}$-fields.

It is obvious that repeating the above path integral calculation
procedure in quadratic approximation we   obtain for transition
probability the same expression (\ref{prob}).

Up to here we have considered only transition from false vacuum to true
vacuum. Now we return to the principal subject of our calculations: the
probability of transition false vacuum+classical field  
$\rightarrow$ true vacuum. The field configuration $A_{K\bar{K}}$ 
which is degenerate with $A_i=A_1=A_c$ can be approximated by
\begin{equation}
A_{K\bar{K}}^c=A_{K\bar{K}}(1+A_c)
\end{equation}
where $A_{K\bar{K}}$ is given by formula (\ref{akk}).

In with  the choice of $A_{K\bar{K}}^c$ in this form we must mention two
facts. First, this approximation for $A_{K\bar{K}}$ implies that we have
fixed the variations of the form of $K$ and $\bar{K}$ beforehand, both in
view of the effect of small classical field and process of motion of $K$
and $\bar{K}$. In the following in the thin wall kink approximation we 
neglect in general the change of the form of $K$ and $\bar{K}$ (the 
corresponding corrections are small). Second, the motion of $K$ and 
$\bar{K}$ now is asymmetric relative to the point $x_0$, since, generally 
speaking, $\Theta(x,t)$ depends in a nontrivial way both on coordinates 
and time. Now proceeding as in Ref. [2] in the thin wall limit 
for the effective Lagrangian we obtain
\begin{equation}
L_{eff}^c  =L_{eff} -\sum_n {a_n}^2
\int_{ x_0-\Lambda_{K}}^{x_0+\Lambda_{\bar{K}}} dx F_n(x,t),
\end{equation}
where $L_{eff}$ is the effective Lagrangian (\ref{lag}) of with
$A_c=0$, and second term in right hand side is the contribution to the
volume energy of the region between $K$ and $\bar{K}$ from small
classical oscillations of the field, and 
\begin{equation}
F_n(x,t)=(1/2)[(\partial_{\mu}\Theta_n)^2-{\kappa_n}^2{\Theta_n}^2]
\end{equation}
is the Lagrangian of  $n$-th normal mode of oscillations.

To calculate the  effective action in the quadratic approximation we expand
$L_{eff}^c$ around new MPEP, that is, classical solution of equation of
motion for $L_{eff}^c$ for imaginary time $\tau$. Proceeding as in Ref.[2]
after simple but slightly cumbersome calculations we obtain up to terms 
$\sim a^2$
\begin{equation}
S^c=S_0^c - (\rho/2)\int_{0}^{\pi} d\theta H,\label{action}
\end{equation}
where
\begin{eqnarray}
&&H_0(\sigma,\tau)=(\partial\sigma/\partial\theta)^2;\\
&&H_1(\sigma,\theta,x_0,t_0) = \nonumber \\
&& [-(\partial \sigma / \partial \theta)^2+
+\sigma^2\sin{\theta}(\partial/{\partial x_0})] Re F(X_0,T_0),\\
&&X_0=x_0+(M/\rho) \sin{\theta} \cos{\theta}/|\cos{\theta}|,\\
&&T_0=t_0 + i(M/\rho)cos{\theta}.
\end{eqnarray}

$S_0^c$ is the "classical" 
action  for MPEP [2]:
\begin{equation}
S_0^c = (\pi M^2/2\rho)(1+(a^2/\rho)D_1(x_0,t_0))
\end{equation}
and the function $D_1$ is defined by the integrals
\begin{eqnarray}
&&D_1(x_0,t_0)=\nonumber\\
&&(2/\pi)Re[\int_{0}^{1} dz z(1-z^2)^{-1/2}[F_n(X_{+}(z);T(z))+
F_n(X_{-}(z);T(z))] \nonumber\\
&&+ \int_{0}^{1} dz\int_{-(1-z^2)^{1/2}}^{(1-z^2)^{1/2}} dz'
F_n(Y(z'),T(z)) - \int_{-1}^{1} dz F_n(Y(z),t_0)].
\end{eqnarray}
where
\begin{eqnarray}
&&X_{\pm}(z) = x_0 \pm (M/\rho)(1-z^2)^{1/2},\\
&&T(z) = i(M/\rho)z+t_0,\\
&&Y(z) = x_0+(M/\rho)z.
\end{eqnarray}

Thus the probability of tunneling transition has the form
\begin{equation}
P_c=(\rho/2\pi \Delta) exp{2S_0^c}\label{prob2}
\end{equation}
where $\Delta=\det'H_0/\det'(H_0+ a^2H_1/\rho)$ and $\det'$ denotes 
a functional determinant with vanishing eigenvalues removed.

Using generalized $\zeta$-function method [18,20] we obtain for 
$\Delta$ the expression
\begin{equation}
\Delta = \exp (a^2 D_2(x_0,t_0)/\rho),\label{del}
\end{equation}
where
\begin{eqnarray}
&&D_2(x_0, t_0) = \rho a^{-2}(d/ds) \zeta_c (s = 0),\\
&&\zeta_c (s) = \sum [(n^2 + a^2 U_n (x_0, t_0) /\rho)^{-s} - n^{-2s}]
\approx (a^2/\rho) \sum U_n n^{-2s-1},\\
&&U_n(x_0, t_0) = Re  \int_{0}^{\pi}d/\theta H_1((2/\pi)^{1/2}
\sin n \theta , \theta, x_0, t_0).
\end{eqnarray}

Substituting (\ref{del}) in the (\ref{prob2}) we obtain
\begin{equation}
P_c=(\rho/2\pi)exp(-\pi M^2/\rho+a^2D(x_0,t_0)/\rho),
\end{equation}
where
\begin{equation}
D(x_0,t_0)=(\pi M^2/\rho)D_1(x_0,t_0)+D_2(x_0,t_0).\label{dae}
\end{equation}

This is the expression for tunneling probability of 
transition from false vacuum + small classical field 
into true vacuum. 
From (\ref{dae}) we see that the presence of small classical field 
in initial state does not change the general form of vacuum tunneling 
probability. The only influence is an appearance of the new factor 
which depends on the classical field and makes the resulting 
probability to be a function of point of the transition origin. 
We do not enter here into a discussion of physical effects
arising from small classical field (see discussion in Ref.[2]) and only
note that the field appearance can stimulate a "catastrophical" (with
$P_c\approx1$) transitions.

\section{The probability of color tube fission}
\bigskip
\indent The main purpose of this section is to obtain the probability $P$ 
of color tube fission including  the small surface oscillation 
effects with pre-exponential factor accuracy. In other words, we want 
to calculate the fission probability $P$ per unit length and per 
unit time of a color tube which  has the radius $r \neq {r_1}$ and 
color  electric field $E=E_1$ and length  $\Delta_1$ at the
moment of its formation, where $r_1$ and $E_1$ are the  stationary color
tube radius and electric field respectively.

   In order to calculate $P$ we use the previously constructed
quasi-phenomenological model [1], based on  (1+1)-dimensional theory of two
scalar fields  $r$  and  $E$  describing the simultaneous change 
of the radius of a tube and a chromoelectric field.

   To study the tube fission it is convenient to choose field variables
$\phi=(4\pi\alpha)^{-1/2}E$ and $\chi=2\log(r/r_0)$ (where 
$\alpha=g^2/4\pi$ is the color interaction constant) .Then the 
Lagrangian of the tube has the form [1]
\begin{eqnarray} 
L=\lambda_2[(\pi/2)(\partial_{\nu}{\phi})^2
+(\alpha/2){\phi}_2(\partial_{\nu}{\chi})^2-\\
-(\epsilon_2^2)(\exp{\chi}+\phi^2\exp{-\chi}) +\cos{2\pi\phi}-1]\nonumber
\end{eqnarray}
where
\begin{eqnarray}
\partial_0=\partial/\partial\tau, 
\partial_1=\partial/\partial\eta,
\end{eqnarray}
and the metric is $(+,-)$.
 
   It should be emphasized that we are dealing with two scalar fields
and, accordingly, the system of vacuum states of a Hamiltonian $H$ 
consists of pairs of fields $(\phi_n,\chi_n)$ that correspond to one 
eigenvalue $H_n$.Here we restrict ourselves to the first false vacuum 
state $(\phi_1,\chi_1)$ that corresponds to a quark tube.

   As stated above in our model  a color tube is kink and antikink 
($K$,$\bar{K}$) that are moving apart in opposite directions and
between which the fields $r$ and $\phi$ are nonzero and equal to 
values$r=r_1+\delta{r}$ and $\phi=\phi_1+\delta\phi$ where $r_1$ 
and $\phi_1$ are the fields that correspond to unstable vacuum and 
$\delta\phi$ and $\delta{r}$ are the small classical fields. The 
breaking of the tube is a transition from the state of pair 
$(K,\bar{K})$ to a state $(K,{\bar{K}}_1, K_1,\bar{K})$ which is
degenerate with it and such that, between extra pair $(\bar{K}_1,K_1)$ 
that are formed inside the tube we have a true vacuum with $\phi_0=0$,
$exp(\chi_0)=0$, $\phi_0\exp(-\chi_0)=0$ (Ref.[2]) 

   Let us consider first the case of a long tube and pair 
$(K_1,\bar{K}_1)$ which is  formed far from  the ends of the tube. 
Then the effect of the ends can be neglected, the tube can be regarded 
as infinite, and one can confine oneself
to the thin wall approximation for $K_1(\bar{K}_1)$.

   As was shown in Ref.[2] the MPEP
connecting initial and final states is the $(\bar{K}_1, K_1)$ 
configuration of fields $(\phi, \chi)$ parameterized by one function
 $\lambda(t)$. Following the
 procedure of Ref. [2] after some calculations we obtain (in the thin wall
approximation) the effective Lagrangian of $\lambda$ in the form [2]
\begin{eqnarray}
 L_{eff}=-
M[(1-{\dot\Lambda_K}^2)^{1/2}+(1-{\dot\Lambda_{\bar{K}}}^2)^{1/2}]+
\rho[\Lambda_{\bar{K}}-\Lambda_K]\\-\sum_{1}^{2}{a_n}^2
\int_{\eta_0-\Lambda_{\bar{K}}}^{\eta_0+\Lambda_K} d\eta 
F^{(n)}(\eta,\tau).\nonumber
\end{eqnarray}
 Here $\rho=\eta^2$ is the energy density in the tube, 
\begin {equation}
M=4\pi^{-1/2}[1+(\alpha/\pi)Z']
\end{equation}
is the renormalized (dimensionless) mass of the soliton, and
\begin{equation}
F^{(n)}=2^{-1}[(\partial_{\nu}\Theta_n)^2-{\kappa_n}^2{\Theta_n}^2]
\end{equation}
is the Lagrangian of small surface oscillations of the tube [2],
\begin{equation}
\dot\Lambda=\partial_0\Lambda.
\end{equation}
and $a_n$ is the amplitude of $n$-th oscillation mode.

Note that the presence of two fields  $\chi$  and $\phi$ in $L_{eff}$ 
manifests itself in the renormalization of the mass of the soliton and in 
the two contributions $\sim{a_n}^2$ corresponding to the two kinds 
of small oscillations [2].
  
   Now  the path integral that define the transition amplitude can be 
calculated in quadratic approximation just in the same way as it was 
made in previous section and one obtains for the tube fission probability 
\begin{equation}
P=(\epsilon^2/2\pi)\exp(-\pi^2M^2/\epsilon^2
+\sum(a_n^2/\epsilon^2)D^{(n)}(\eta_0,\tau_0))
\end{equation}
where term $D(\eta_0,\tau_0)$ accounts for the small oscillations of 
a tube surface that are created at the moment of tube formation.

   The functions $D^{(n)}$ contain the contributions of two different origin
\begin{equation}
D^{(n)}=D_1^{(n)}+D_2^{(n)}.
\end{equation}
The part $D_1^{(n)}$ comes from the classical action for MPEP and has the
form 
\begin{eqnarray}
&&D_1^{(n)}(x_0,t_0)=\nonumber\\
&&(2/\pi)Re[\int_{0}^{1} dzz(1-z^2)^{-1/2}
(F_n(x_0+(M/\rho)(1-z^2)^{1/2};i(M/\rho)z+t_0)+\nonumber\\
&&F_n(x_0-(M/\rho)(1-z^2)^{1/2};i(M/\rho)z+t_0))+\\
&&\int_{0}^{1} dz\int_{-(1-z^2)^{1/2}}^{(1-z^2)^{1/2}} dz'F_n(x_0+(M/\rho)z',
i(M/\rho)z+t_0)-\nonumber\\
&&\int_{-1}^{1} dzF_n(x_0+(M/\rho)z,t_0)].\nonumber
\end{eqnarray}

The second contribution $D_2^{(n)}$ comes from the determinant of
operator that describe the quadratic corrections to MPEP action in path
integral. For $D_2^{(n)}$ one obtains (see sec.2)
\begin{equation}
D_2^{(n)}=(d/dq)\zeta_c^{(n)}(q)|_{q=0}
\end{equation}
where
\begin{equation}
\zeta_c^{(n)}(q)=\sum_kV_{kk}(\eta_0,\tau_0)k^{-2q-1},
\end{equation}
and
\begin{eqnarray}
&&V_{kk}(\eta_0,\tau_0)=\nonumber\\
&&Re\int _{0}^{\pi} d\theta[-(2/\pi)k^2\cos^2k\theta+
\sin^2k\theta\sin\theta\partial_0] \\
&&F^{(n)}(\tau_0+
(M/\epsilon^2)\sin\theta\cos\theta/|\cos\theta|,
i(M/\epsilon^2)\cos\theta+\tau_0).\nonumber
\end{eqnarray}

   As noted above the large-mass oscillations $(\Theta_1)$ cause fission
of tube in a very short time. Thus the
construction of tube itself terminates after all high frequency 
oscillations have been "used" on this fission and a tube has been 
formed in which only low-frequency interrelated oscillations of 
a radius and of a electric field $\delta r/r_1\sigma_2\Theta_2$ 
and $\delta\phi/\phi_1\sigma_2\Theta_2$ are kept. This means that 
on the tube which is formed, only those amplitudes of the oscillations 
of small mass $\kappa_2$ survive for which the probability of
fission stimulated by these oscillations can no longer become $\sim1$. 
It can be easily shown [2] that for sufficiently large $\tau_0$ and 
far from the edges of the tube with $r=r_1+\delta r$ and 
$\phi=\phi_1$ at the moment of tube formation $\tau_0=0$, the field 
$\Theta_2$ is the function of $s=\tau_0^2-\eta_0^2$  only. Accordingly, 
$D^{(2)}$ is the function of s and one have for probability of a 
tube fission 
\begin{equation}
P=(\epsilon^2/2\pi)\exp (-\pi 
M^2/\epsilon^2+(a_2^2/\epsilon^2)D^{(2)}(s)).\label{prob3}
\end{equation}

   So far we have considered one MPEP soliton or,equivalently, tube 
fission via quark-antiquark pair of one type creation inside the tube.
 In real world the tube fission take place via $q\bar{q}$ pairs of 
many flavors or various diquark ect.pairs creation. In our tube model 
to each type of particle-antiparticle pair corresponds its own MPEP or, 
other words, its own $(K,\bar{K})$-soliton with mass $M_i$. This means 
that in the functional integral contribute many different MPEP's and 
the total probability $P$ is the sum of partial
probabilities $P_i$ of each type MPEP's. Thus one obtains
\begin{eqnarray}
&&P=\sum P_i,\label{prob4}\\
&&P_i=(\epsilon^2/2\pi)\exp(-\pi M_i^2/\epsilon^2
+ (a_2^2/\epsilon^2)D_i^{(2)}(s)).\label{prob5}
\end{eqnarray}

It is evident that the sum (\ref{prob4}) is dominated by the term 
of smallest kink mass. However to calculate the inclusive spectra 
of primary hadrons of different types we need also other partial 
probabilities.

   When the distance of the kink wall from the end of the tube 
becomes of the order of the thickness of the wall, the influence 
of the end of the tube
becomes important and, strictly speaking, the thin wall approximation 
for the$K_1(\bar{K}_1)$ that has been formed near the end of the tube 
is no longer applicable. It was shown in Ref.[3] that the mass of $K_1$ 
increases with the decrease of the distance between the world point of 
the center of fission $(\tau_0,\eta_0)$ and the end of the tube,  the 
probability of the fission should decrease near of 
the tube end. Thus, even if the previous fission occurred at a distance 
$l=M/\epsilon^2$ from the tube end, first the fragment that has been 
formed "expands" in such a way that its length become greater 
than $2M/\epsilon^2$ and only then does the real chance for 
the tube to break again.

   \section{The probability of tube fission at $n$ points}
\bigskip

\indent In our model the tube is formed by the color electric field of the 
moving (in opposite directions in $CM$ frame )  $K$ and $\bar{K}$. 
The motion of $K$ and $\bar{K}$ is decelerated by the tube tension 
and they will oscillate back and forth just as the yo-yo relativistic 
string. After some time the tube will break into two parts by production 
of a pair of $K$ and $\bar{K}$ at the space-time point $(\tau_1, \eta_1)$.
 At later time another pair $K,\bar{K}$ will be produced at 
$(\tau_2, \eta_2)$ and so on.The successive breaking of  the tube 
form pieces of small length that are primary hadrons.

For further discussion it is 
useful to introduce light-cone variables $u=\tau - \eta$ and 
$v=\tau + \eta$ and consider the process of
tube breaking in $(u,v)$ coordinates in CM frame.

We begin with the case of one type of kink. Let us consider first the
probability of tube fission into two pieces i.e. let the $dP(1)$ be the
probability of tube breaking in only one world point. Any
breaking in the point $(\tau_1, \eta_1)$ means that now we have two
independently developing tubes. The corresponding probability is the
product of two factors: one is the probability $\exp[-W(S_1)]$ that 
there is not any breaking before the world point $(1) = (u_1,v_1)$ or, 
in other words, inside the area $S_1$, limited by the continuation 
of trajectories of created $K$ and $\bar{K}$ ($S_1$ is the area 
$(0,1', 1, 1")$   on Fig.1
or is the strip $0\leq u'\leq u_1, 0\leq v'\leq v_1)$ where
\begin{equation}
W(S_1)= (1/2) \int_{S_1}du'dv'P((u'+ v')/2; (u'-v')/2).\label{prob6}
\end{equation}
The second factor 
\begin{equation} 
dw(1) = (1/2)P((u_1+v_1)/2; (u_1-v_1)/2)du_1dv_1\label{prob7}
\end{equation}  
is the probability of tube breaking in the point $(1)$.

Thus we have
\begin{equation}
dP(1) = dw(1) e^{-W(1)}
\end{equation}

Now we pass to the fission at the points
$(u_1,v_1)=(1)$ and $(u_2,v_2)=(2)$ (see fig.2). There are two
different cases. The first one is that there are supplementary 
breaking between points $(1)$ and $(2)$ (or in the area $(2,D,1,C)$). 
Then the  probability $dP(2,0)$ is the product of :\\
(i) the probability $dP(1)$ that the first breaking point is $(1)$;\\ 
(ii) the probability $d \bar{P}(2) = dw(2)e^{W(S_2)}$ ,
(where $S_2$ is the area $(v_1, C,2, v_2)$), that the
point  $(2)$ is just the new breaking point [5,10].  

Thus we have
\begin{equation}
dP(1,2) = dP(1) d\bar{P}(2) = dw(1)dw(2) e^{-W(S)},
\end{equation}
where $S$ is the area $(0,u_1, 1, C, 2, v_2)$ or sum of strips $0\leq 
u'\leq u_1, 0\leq v'\leq v_1$ and $0\leq u'\leq u_2, v_1\leq v'\leq v_2$.

If the piece produced due the breaking $(1)$ and $(2)$ is not broken
and develops like the "yo-yo" string then, we have an additional factor 
$\exp[-W(S_3)]$ where $S_3$ is the area $(C,1,D,2)$, that is, the
probability that  there is no any breaking in the $(C, 1, D, 2)$.
Finally we have the same expression (70) with the only replacement 
$S\rightarrow \bar{S} = (0,1, D, 2) = [0\leq u'\leq u_1, 0\leq v'\leq v_2]$.

Now it is easy to see that the probability of n successive breaking
can be written in the form
\begin{equation}
dP(1,2,...,n) = [\prod_{i=1}^{n}dw(i)]e^{ - W(S_n)},
\end{equation}
where the integration area is limited  by the continuation  of
trajectories of created $K$ and $\bar{K}$-s. For small kink mass the
integration area $S_n$ is the sum of successive strips (see fig.3)
\begin{eqnarray} 
&&[0\leq u'\leq u_1, 0\leq v'\leq v_1;  0\leq u'\leq u_2,\nonumber \\
&&v_1\leq v'\leq v_2;...; 0\leq u'\leq u_n, v_{n-1}\leq v'\leq v_n] 
\end{eqnarray}

For each stable piece produced by $i$ -th and $i+1$ -th adjacent
breaking the supplementary strip $[u_{i=1}\leq u'\leq u_i,
v_i\leq v'\leq v_{i+1}]$ must be added.

For many types of kinks repeating the above arguments we obtain for the
probability of $n$ successive breaking when at point $(1)$ is created
$K\bar{K}$ - pair of type $a_1$, at point $(2)$  - of type $a_2$  etc. the
following expression 
\begin{equation}
dP(1,a_1; 2,a_2; ..., n,a_n) = [\prod _{i=1}^{n} dw(i,a_i)]e^{-W_t},
\end{equation}
where $dw(i,a_i)$is the partial  probability of tube breaking at the
point $(i)$ due the creation $K\bar{K}$ pair of type $a_i$ given by the
expression ~(\ref{prob6}) and $W_t$ is calculated using the formula 
~(\ref{prob7}) with evident
replacement $p\rightarrow$ [total probability] $= p_t$ with  $p_t$
given by Eq. (\ref{prob4}).

\section{Conclusion and outlook}
\bigskip

\indent The above expressions enable us to  calculate the exclusive and
inclusive spectra of primary hadrons, i.e., the spectra of pieces of given 
length. The procedure of calculation is the following. We must 
integrate the probability of fission that form the pieces  with 
 momenta $p_1, p_2,..., p_n$ over all allowed 
positions of fission space-time points. It is convenient to use the the 
light-cone coordinates $u$ and $v$ and light-cone momenta $p_{\pm,j} = 
p_j \pm E_j$ and center-of-mass system of parent tube. Since the 
light-cone momentum of piece produced by pair $(K_j, \bar{K})$ is given by 
\begin{eqnarray}
	 && p_{+ jl} = \epsilon^2(u_j - u_l)\\
	 && p_{- jl} = \epsilon^2(v_j - v_l) 
\end{eqnarray}
the requirement that produced piece has  light-cone  momentum 
equal to $p_{\pm,j.l}$ means that the difference $\delta u_{j,l}$ must be 
equal to  $\tilde{p}_{j,l} = p_{+,j,l}/ \epsilon^2$. We can take this into 
account by introducing into integrand  the  $\delta (u_j 
- u_l - \tilde{p}_{j,l})$  and $\delta (v_j - v_l - 
m^2_{jl,t}/\tilde{p}_{j,l})$, where  $m^2_{jl,t}$ is the piece 
transversal mass, for each piece. 
Then we choose such an ordering of the fission points that corresponds to 
the momentum space region we are interested in. (It must be noted  that each 
ordering of fission points corresponds to its specific region of momentum 
space.) The fission point ordering can be easily taken into account by 
introducing into the integrand $\theta$ - functions for $u$ and $v$ 
coordinates. Now the integration area is the $2n$-cube ($0\leq u_1 \leq 
\tilde{P}, 0 \leq_0 v_1 \leq \tilde{P}_0;..., 0 \leq u_n \leq 
\tilde{P}_0, 0 \leq v_n \leq \tilde{P}_0$) where $n$ is the number of 
breaking points.

For exclusive spectra the integrand is the probability of $n-1$ adjacent 
fission only. This means that all pieces (except two that contain 
the edges of parent tube) have the common point of $(K,\bar{K})$ pair creation.

Rather more complicated situation is for an inclusive spectra.
In this case we have distinct contributions that correspond to 
particles (pieces) of different adjacent ranges on the Field-Feynman 
terminology [22]. 
 The $k$-particle inclusive spectra of 
primary hadrons are the sum of probabilities:\\
 (i) $P_{2k}$ of the tube breaking in the $2k$ different points,\\
 (ii)$P_{2k-1}$ of the tube breaking in the $2k-1$ different points 
 when one wall of the piece is the edge of parent tube,\\
 (iii)$P_{2k-2}$ of the tube breaking in the $2k-2$ different points 
 when there are two pieces the one wall of each is the edge 
 of parent tube.
Thus the inclusive spectra are defined by integrals over the 
probability of at least $m (m \geq  n+1)$ fission independent on there 
are or not any other fission. The probability of 
creation of pieces with common $(K,\bar{K})$ pairs to IS contribute as
 well as the cases 
when all or part of $(K,\bar{K})$ of adjacent pieces belong to different 
$(K,\bar{K})$ pairs (i.e. between these $K$ and $\bar{K}$  are any 
supplementary  fission points).

{\bf REFERENCES}
\bigskip
\begin{itemize}
\item[1] Abramovskii V.A., E.V. Gedalin ,  E.G. Gurvich
 and O.V.Kancheli O.V. Yad.Fiz. {\bf 48}, 1805
(1988) [Sov. J. Nucl. Phys. {\bf 48},1086 (1988)].
\item[2] Abramovskii V.A., E.V.Gedalin, E.G.Gurvich 
 and O.V.Kancheli. Yad. Fyz.
{\bf49}, 1741(1989) [Sov. J. Nucl. Phys.{\bf 49}, 
1078, (1989)].
\item[3]Gedalin E.V. and E.G.Gurvich.  Yad.Fiz. 
{\bf52}, 240, (1990) [Sov. J. Nucl. Phys.
{\bf52} 153(1990)].
\item[4]Gedalin E.V.Phys. Rev.{\bf D48}, 1424 (1993).
\item[5] Artru X. and G,Menessier. Nucl.Phys.{\bf B70},93(1974);
Artru X.Phys.Rep.{\bf 97},147(1983)
\item[6]Gottschalk T.D. and D.A.Morris, Nucl. Phys.{\bf B288}, 729 
(1987);
Morris D.A., Nucl. Phys.{\bf B288}, 717 
(1987).
\item[7] Fong C.P. and B.R.Webber, Nucl. Phys.{\bf B355}, 54(1991).
\item[8] Andersson B. and P.A.Henning, Nucl.Phys. {\bf B355}, 82(1991).
\item[9] Gustavson G. and A.Nilson, Nucl.Phys. {\bf B355}, 
106(1991).
\item[10] Andersson B., G.Gustavson, G.S.Ingelman, 
and T.Sjostrand,\   Phys.Rep.{\bf 97C}, 31(1983) and references therein.
\item[11] Werner K. Phys. Rep. {\bf 232} 87 (1993) and references therein.
 \item[12] Bitar K. and S.-J.Cheng,\  Phys. Rev.{\bf D12} 489 (1978).
 \item[13] Aflec I. and F.De Luccia, Phys.Rev.{\bf D20} 3168 (1979).
\item[14] Casher A., H.Neuberger, S.Nussinov, Phys. Rev. {\bf D20}, 179
(1979).
\item[15] Coleman S. Phys. Rev.{\bf D15}, 2929(1977);"The Uses of 
Instantons" in The Ways of Subnuclear Physics, Plenum pub. NY,1979.
\item[16] B.De Witt. Phys.Rep. {\bf 50}, 255 (1979).
\item[17] Gouncharov A.S. and A.D.Linde, \ Fiz. Elem. Chastitz At. Yadra 
{\bf 17} 837 (1986)  [Sov. J. Part. Nucl. {\bf 17} 369 (1986)]
and references therein.
\item[18] Ramon P. "Field Theory. A modern premier." Benjamin, New York, 
1981.
\item[19] De Witt B. General Gravity. An Einstein centenary survey ed. by 
S.W.Hawking and S.Israel. Cambridge Univ. Press. Cambridge, !979.
\item[20] Hawking S.W.  General Gravity. An Einstein centenary survey ed. by 
Hawking S.W. and S.Israel. Cambridge Univ. Press. Cambridge, !979.
\item[21] Voloshin M.B.Yad.Fiz. {\bf  42}, 1012 (1985); {\bf 43} 769 
(1985).
[Sov. J . Nucl. Phys. {\bf 42} 644 (1985), {\bf 43} 488 (1986)].
\item[22] Field R. and R.P.Feynman. \  Nucl. Phys.{\bf B 123} 429 (1977).

\end{itemize}

\pagebreak

{\bf FIGURE CAPTIONS}
\bigskip
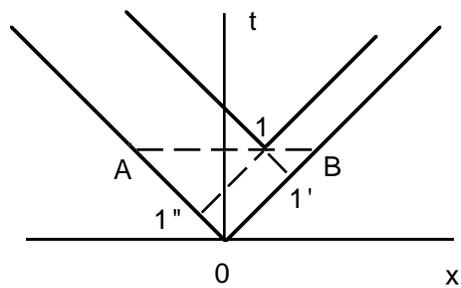
Fig.1. The one breaking point world diagram in ($t,x$) and ($u,v$) 
coordinates.

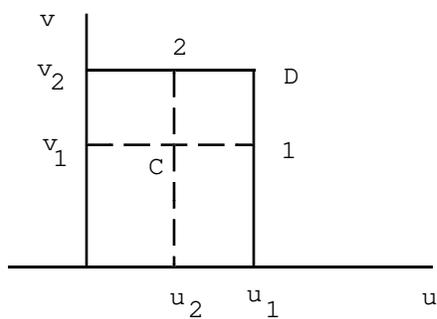
Fig.2.The two breaking of the tube in world points (1) = ($u_1,v_1$) 
and (2) = ($u_2,v_2$).

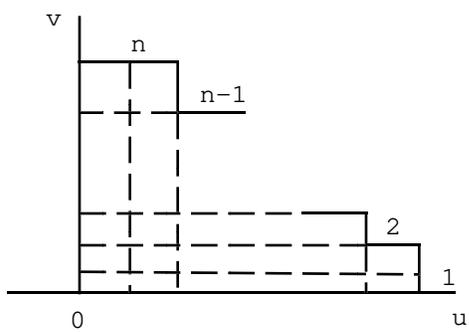
Fig.3. The integration area for $n$ breaking of the tube.

\end{document}